\documentclass[%
prl,aps,amsmath,floatfix,superscriptaddress,showpacs,twocolumn%
]{revtex4}
\usepackage{bm}
\usepackage{amsmath}
\usepackage{graphicx}
\begin{document}
\title{Temperature dependence of rho- and $\bm{a_1}$-meson masses\\
and mixing of vector and axial-vector correlators}
\author{Michael Urban}
\affiliation{Institut de Physique Nucl{\'e}aire, F-91406 Orsay Cedex,
France}
\author{Michael Buballa}
\affiliation{Institut f{\"u}r Kernphysik, Schlossgartenstr.\ 9,
D-64289 Darmstadt, Germany}
\affiliation{Gesellschaft f{\"u}r Schwerionenforschung, Planckstr.\ 1,
D-64291 Darmstadt, Germany}
\author{Jochen Wambach}
\affiliation{Institut f{\"u}r Kernphysik, Schlossgartenstr.\ 9,
D-64289 Darmstadt, Germany}
\begin{abstract}
Within a chiral model which provides a good description of
the properties of $\rho$ and $a_1$ mesons in vacuum, it is shown that,
to order $T^2$, the $\rho$- and $a_1$-meson masses remain constant in
the chiral limit, even if at tree level they are proportional to the chiral 
condensate, $\sigma_0$. Numerically, the temperature dependence of
the masses turns out to be small also for realistic parameter sets and
high temperatures. The weak temperature dependence of the masses
is consistent with the Dey-Eletsky-Ioffe mixing theorem, and
traces of mixing effects can be seen in the spectral function
of the vector correlator at finite temperature.
\end{abstract}
\maketitle

It is commonly believed that the enhancement of the dilepton spectrum
in the invariant mass range around $400$ MeV observed in heavy-ion
experiments \cite{CERES} is a consequence of medium modifications of
the $\rho$-meson spectral function, i.e., more precisely, the spectral
function of the vector-isovector current-current correlation function,
in hot and dense matter. Most of the more ``conventional''
calculations find a strongly increasing $\rho$ meson width (see Ref.\
\cite{RappWambach} and references therein), but the more speculative
hypothesis of a ``dropping $\rho$ mass'' by Brown and Rho
\cite{BrownRho} is being discussed as a viable alternative
\cite{BrownRho2}. The aim of the present study is to combine the
phenomenology included in the more ``conventional'' calculations with
the constraints imposed by chiral symmetry. As a first step, we have
constructed in Ref.~\cite{UBW} a chiral model which gives a good
description of the vector and axial-vector spectral functions in
vacuum. The model is based on the linear $\sigma$ model, which has
been extended to incorporate the $\rho$ meson and its chiral partner,
the $a_1(1260)$, as elementary fields. In this letter we will discuss
results from the application of this model to finite temperatures.

Within the model presented in Ref.\ \cite{UBW} the vector meson masses
at tree level are given by
\begin{equation}
m_\rho^2 = m_0^2+h_2 \sigma_0^2\,,\quad
m_{a_1}^2 = m_0^2+(h_1+h_2)\sigma_0^2\,,
\label{mtree}
\end{equation}
where $m_0$ is the vector-meson mass parameter in the Lagrangian,
$h_1$ and $h_2$ are coupling constants, and $\sigma_0$ is the
expectation value of the scalar field $\sigma$ at tree level. As shown
in Ref.\ \cite{UBW}, the parameter $m_0$ can take any value below
$\approx 400$ MeV, in particular, $m_0 = 0$. Thus, if Eq.\
(\ref{mtree}) were valid in general, i.e., not only at tree level,
``Brown-Rho scaling'' would emerge, i.e., the vector meson masses
would go to zero at the chiral phase transition, where
$\langle\sigma\rangle$ vanishes.

A good description of the experimentally measured vector and
axial-vector spectral functions in vacuum \cite{ALEPH} can be achieved
by including the one-loop $\rho$ and $a_1$ self-energies, which
comprise the dominant decay channels $\rho\rightarrow\pi\pi$ and
$a_1\rightarrow\pi\rho$ as well as many other one-loop graphs which
are necessary to comply with chiral symmetry \cite{UBW}. The
divergences generated by the loops can be canceled by chirally
symmetric counterterms, which are adjusted such that masses at one
loop are equal to those at tree level \cite{UBW}. However, when we
calculate the same diagrams at finite temperature $T$ (i.e., when we
replace the loop integrals over $k_4$ by a summation over discrete
Matsubara frequencies $2\pi n T$), the masses at one loop become
temperature dependent, while the masses at tree level remain constant.

\begin{figure}
  \includegraphics[bb=69 660 532 760, width=8.5cm]{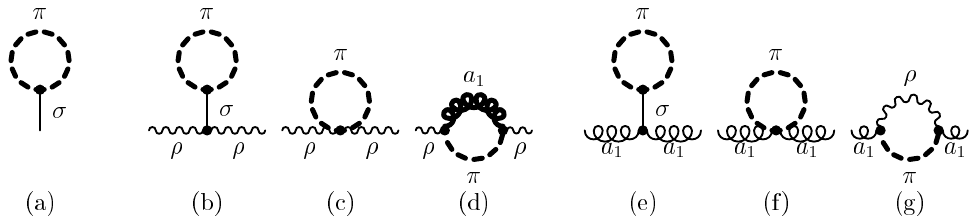}
\caption{Diagrams generating the dominant medium contributions to
$\langle\sigma\rangle$ (a), $\Sigma_\rho$ (b--d) and
$\Sigma_{a_1}$ (e--g).
\label{diagSigma}}
\end{figure}

Let us start by inspecting the temperature dependence of the
condensate $\langle\sigma\rangle$. In the one-loop approximation it
can be written as
\begin{equation}
\langle\sigma\rangle = \sigma_0+\Delta\sigma_0\,,
\end{equation}
where $\Delta\sigma_0$ denotes the sum of all one-loop tadpole graphs,
divided by $i m_\sigma^2$ \cite{UBW}. It can be decomposed into
vacuum and medium contributions, $\Delta\sigma_0 =
\Delta\sigma_{0\,\mathit{vac}} + \Delta\sigma_{0\,\mathit{med}}$. The
dominant medium contribution is generated by diagram (a) shown in
Fig.\ \ref{diagSigma}.
In the chiral limit, $m_\pi\to 0$, it is of the order $T^2$, while all
other one-loop diagrams are either vanishing ($\propto m_\pi^2$) or
suppressed by $e^{-m/T}$, if the particle propagating in the loop has
mass $m$. The explicit expression corresponding to diagram (a) reads
\begin{equation}
\Delta\sigma_{0\,\mathit{med}}^{\text{(a)}} =
  -\frac{3\sigma_0\lambda^2 Z_\pi}{2\pi^2 m_\sigma^2}
  \!\!\int_0^\infty\!\!\!\!d|\bm{k}|\,\bm{k}^2\frac{n_\pi}{\omega_\pi}\,,
\label{Tadpole}
\end{equation}
where the abbreviations $\omega_\pi = \sqrt{\bm{k}^2+m_\pi^2}$ and
$n_\pi = 1/(e^{\omega_\pi/T}-1)$ have been used. The symbol $\lambda$
denotes the usual coupling constant of the linear $\sigma$ model, and
$Z_\pi = Z_{\pi\,1}$ is the residue of the pion propagator at tree
level \cite{UBW}.

In the numerical calculations, the finite parts of the counterterms
are determined such that $f_\pi$ in vacuum remains unchanged by the
loop corrections. In the chiral limit this has the consequence that
$\Delta\sigma_{0\,\mathit{vac}}$ vanishes. Hence, we can replace
$\sigma_0$ in Eq.\ (\ref{Tadpole}) by
$\langle\sigma\rangle_{T=0}$. For further simplification we make use
of the relations $m_\sigma^2 = 2\lambda^2\sigma_0^2$ and $f_\pi =
\sigma_0/\sqrt{Z_\pi}$, which follow from those given in Ref.\
\cite{UBW} in the limit $m_\pi = 0$. In this way we finally obtain
\begin{equation}
\langle\sigma\rangle_{T\neq 0}
  = \langle\sigma\rangle_{T=0}\Big(1-\frac{T^2}{8 f_\pi^2}+\dots\Big)\,,
\label{sigma0T2}
\end{equation}
as required by chiral symmetry \cite{GasserLeutwyler}.

Let us now turn to the temperature dependence of the $\rho$
self-energy $\Sigma_\rho(q_0,\bm{q})$, where $q_0$ and $\bm{q}$ denote
the energy and three-momentum of the $\rho$ meson relative to the heat
bath. In the following we will restrict ourselves to $\bm{q} = 0$. In
this case the self-energy tensor $\Sigma_\rho^{\mu\nu}$ has, as in
vacuum, only two independent components, namely the four-dimensionally
longitudinal and transverse self-energies, $\Sigma_\rho^l$ and
$\Sigma_\rho^t$. In what follows we will concentrate on the latter and
denote it by $\Sigma_\rho$ for simplicity. Similar to
$\Delta\sigma_0$, the self-energy $\Sigma_\rho$ can be decomposed into
vacuum and medium contributions. The dominant medium contributions are
generated by the diagrams (b) to (d). The two-pion loop, which is the
dominant graph in vacuum, does not contribute to order $T^2$, because
the pions couple to the $\rho$ meson in $p$-wave, resulting in
additional powers of the thermal pion momentum. The explicit
expressions for the three contributions lead to
\begin{align}
\Sigma_{\rho\,\mathit{med}}^{\text{(b)}}
 &= 2 h_2 \sigma_0 \Delta\sigma_{0\,\mathit{med}}
\label{Sigmarhob}\\
\Sigma_{\rho\,\mathit{med}}^{\text{(c)}}
 &= \frac{(2 h_1+3 h_2) Z_\pi}{2 \pi^2}
    \!\!\int_0^\infty\!\!\! d|\bm{k}|\bm{k}^2
    \frac{n_\pi}{\omega_\pi}\,,
\label{Sigmarhoc}\\
\Sigma_{\rho\,\mathit{med}}^{\text{(d)}}
 &= \frac{h_1^2\sigma_0^2 Z_\pi}{2 \pi^2}
    \!\!\int_0^\infty\!\!\! d|\bm{k}|\bm{k}^2
    \frac{n_\pi}{\omega_\pi\omega_{a_1}}
\nonumber\\
 &\quad\times \Big(
    \frac{\omega_{a_1}-\omega_\pi}{q_0^2-(\omega_{a_1}-\omega_\pi)^2}+
    \frac{\omega_{a_1}+\omega_\pi}{q_0^2-(\omega_{a_1}+\omega_\pi)^2}
    \Big)\,.
\label{Sigmarhod}
\end{align}
(To obtain real and imaginary part, $q_0$ must be replaced by
$q_0+i\varepsilon$.) The contribution of diagram (d) takes a much
simpler form in the chiral limit if, in addition,
$T\ll|q_0-m_{a_1}|$. In this case, to order $T^2$, the pion momentum
$\bm{k}$ can be neglected in the second line of Eq.\
(\ref{Sigmarhod}), i.e., $\omega_\pi = 0$ and $\omega_{a_1} =
m_{a_1}$, and the sum of the three contributions reads
\begin{equation}
\Sigma_{\rho\,\mathit{med}}(q_0,0) = \frac{T^2 h_1 \sigma_0^2}{6f_\pi^2}\,
  \frac{q_0^2-m_\rho^2}{q_0^2-m_{a_1}^2}\,.
\label{Sigmarho}
\end{equation}
The change of the $\rho$ mass is governed by the real part of the
self-energy at $q_0\approx m_\rho$, but Eq.\ (\ref{Sigmarho}) implies
in particular $\Sigma_{\rho\,\mathit{med}}(m_\rho,0) = 0$. Thus, to
order $T^2$, the $\rho$ mass does not change at all, even if the
condensate decreases, because the self-energy contribution
$\propto\Delta\sigma_{0\,\mathit{med}}$, diagram (b), is exactly
canceled by the other diagrams, (c) and (d).

For the $a_1$ meson the situation is analogous. The dominant medium
contributions to the $a_1$ self-energy are given by the diagrams (e)
to (g). In exactly the same way as described above for $\Sigma_\rho$,
one finds, to order $T^2$:
\begin{equation}
\Sigma_{a_1\,\mathit{med}}(q_0,0) = -\frac{T^2 h_1 \sigma_0^2}{6f_\pi^2}\,
  \frac{q_0^2-m_{a_1}^2}{q_0^2-m_\rho^2}.
\label{Sigmaa1}
\end{equation}
This shows that also the $a_1$ mass remains constant. The fact that
the $\rho$ and $a_1$ masses do not change to order $T^2$ was already
found within the so-called gauged linear $\sigma$ model some time ago
\cite{Pisarski}. However, within the gauged linear $\sigma$ model,
this is less surprising, since the $\rho$ mass is independent of
$\sigma_0$ at tree level, and diagram (b) does not exist.

The results presented above are valid only in the chiral limit and at
low temperatures. The more general case of $m_\pi\neq 0$ and higher
temperatures can be analyzed numerically. Then, of course, not only
the diagrams shown in Fig.\ \ref{diagSigma}, but all one-loop
self-energy diagrams have to be included. We will use two different
parameter sets: one of those listed in Ref.\ \cite{UBW} (parameter set
A), and one with a very small pion mass ($m_\pi = 0.1$ MeV)
appropriate for the chiral limit. In both parameter sets we have
chosen $m_0 = 0$, i.e., naively one would have expected "Brown-Rho
scaling", as explained below Eq.\ (\ref{mtree}).

\begin{figure}
\includegraphics[width=7cm]{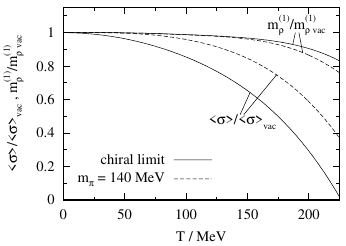}
\caption{The temperature dependence of the condensate
$\langle\sigma\rangle$ and of the $\rho$ mass $m_\rho^{(1)}$ in the
chiral limit (solid lines) and for a realistic parameter set (dashed
lines). All curves are normalized to their values at $T = 0$.
\label{mrhoT}}
\end{figure}

As in Ref.~\cite{UBW}, we define the one-loop corrected $\rho$ mass
$m_\rho^{(1)}$ as the position of the maximum of the $\rho$ spectral
function, rather than as the zero of the real part of the $\rho$
propagator. This implies that $m_\rho^{(1)}$ could in principle
change, even if $\Sigma_{\rho\,\mathit{med}}$ vanishes at
$q_0=m_\rho$. However, this subtlety does not change our conclusions
for the low-temperature behavior of $m_\rho^{(1)}$ significantly, as
can be seen in Fig.\ \ref{mrhoT}.
In the chiral limit, $m_\rho^{(1)}$ stays almost constant at low $T$,
whereas the condensate decreases rapidly according to Eq.\
(\ref{sigma0T2}). At very high temperatures ($T\gtrsim 150$ MeV), of
course, the low-temperature expansion is no longer valid, and also the
$\rho$ mass starts to change. In the case $m_\pi = 140$ MeV, the main
difference to the chiral limit is that all temperature effects are
suppressed by $e^{-m_\pi/T}$, and therefore also the condensate
remains almost constant below $T\approx 50$ MeV. At higher
temperatures, the cancellation of the different contributions to
$\Sigma_{\rho\,\mathit{med}}$ is not as perfect as in the chiral
limit, resulting in a somewhat stronger temperature dependence of
$m_\rho^{(1)}$ than in the case $m_\pi = 0$.

Our results corrobate earlier arguments \cite{RappWambach} that in the
chiral limit a change of the $\rho$ or $a_1$ mass of the order $T^2$
would violate the Dey-Eletsky-Ioffe mixing theorem and is therefore
forbidden by chiral symmetry. This mixing theorem, which relies only
on the current algebra of chiral symmetry, states that to order $T^2$
the vector and axial-vector current-current correlation functions
$\Pi_V(q_0,\bm{q})$ and $\Pi_A(q_0,\bm{q})$, respectively, behave as
follows \cite{DeyEletsky}:
\begin{align}
\Pi_V &=
  (1-\epsilon)\Pi_{V\,\mathit{vac}}+\epsilon\Pi_{A\,\mathit{vac}}\,,\\
\Pi_A &=
  (1-\epsilon)\Pi_{A\,\mathit{vac}}+\epsilon\Pi_{V\,\mathit{vac}}\,,
\label{mixing}
\end{align}
with $\epsilon = T^2/(6f_\pi^2)$.

\begin{figure}
\includegraphics[bb=56 660 540 761, width=8.6cm]{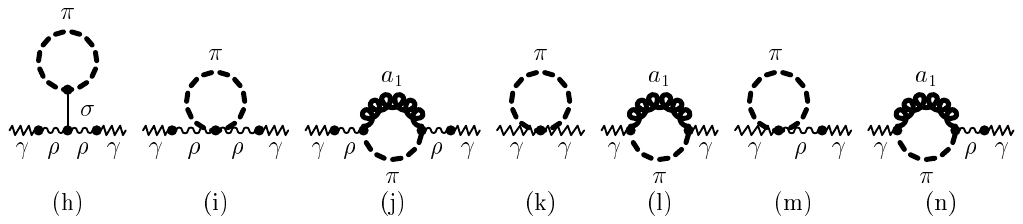}
\caption{Diagrams for the dominant medium contributions to
$\Sigma_\gamma$ in a strict loop expansion to one loop order.
\label{diagPi}}
\end{figure}

It should be noted that the mixing coefficient $\epsilon$ is of order
$\hbar$, since the mixing is generated by thermal pion loops. Hence,
if the correlators are calculated in a strict loop expansion, the
mixing theorem cannot be fulfilled exactly. E.g., if $\Pi^{(n)}_V$ and
$\Pi^{(n)}_A$ denote the correlators to order $\hbar^n$ (i.e., up to
$n$ loops), the medium contribution to $\Pi^{(1)}_V$ must be given by
\begin{equation}
\Pi_{V\,\mathit{med}}^{(1)} =
  \epsilon(\Pi_{A\,\mathit{vac}}^{(0)}-\Pi_{V\,\mathit{vac}}^{(0)}).
\label{mixing1loop}
\end{equation}
Since our Lagrangian is chirally symmetric, this relation is certainly
fulfilled. At tree level the transverse parts of the correlators read
\begin{align}
\Pi_V^{(0)}(q^2) &= -\frac{(f q^2)^2}{q^2-m_\rho^2}\,,\\
\Pi_A^{(0)}(q^2) &= -\frac{(f q^2-g\sigma_0^2)^2}{q^2-m_{a_1}^2}
  -\sigma_0^2\,,
\label{PiA0}
\end{align}
where $f$ and $g$ denote the $\gamma\rho$ and $\rho\pi\pi$ coupling
constants, respectively \cite{UBW}. The pion pole does not appear in
Eq.\ (\ref{PiA0}), because it contributes only to the longitudinal
part of the axial-vector correlator. The last term in Eq.\
(\ref{PiA0}) can be understood as the contribution of the
$WW\sigma\sigma$ seagull vertex, if $\Pi_A$ is computed via the
$W$-boson self-energy. Next we turn to the left-hand side of Eq.\
(\ref{mixing1loop}), i.e., the medium contribution to
$\Pi_V^{(1)}$. To order $T^2$ it is generated by the diagrams shown in
Fig.\ \ref{diagPi}.
(As in Ref.\ \cite{UBW}, we draw diagrams for the photon self-energy
$\Sigma_\gamma = -e^2\Pi_V$ rather than for $\Pi_V$.) The contribution
of the diagrams (h) to (j) can be expressed in terms of the $\rho$
self-energy:
\begin{equation}
\Pi_{V\,\mathit{med}}^{(1\,\text{h--j})}
  = -\frac{(fq_0^2)^2}{(q_0^2-m_\rho^2)^2}\Sigma_{\rho\,\mathit{med}}\,.
\end{equation}
The diagrams (k) to (n), which can be evaluated with the same
techniques as described for $\Sigma_{\rho\,\mathit{med}}$, give
\begin{align}
\Pi_{V\,\mathit{med}}^{(1\,\text{k})} &= -\frac{T^2\sigma_0^2}{6f_\pi^2}\,,\\
\Pi_{V\,\mathit{med}}^{(1\,\text{l})} &=
  -\frac{T^2 g^2\sigma_0^4}{6f_\pi^2}\,\frac{1}{q_0^2-m_{a_1}^2}\,,\\
\Pi_{V\,\mathit{med}}^{(1\,\text{m})} &=
  \frac{T^2 g\sigma_0^2 fq_0^2}{6f_\pi^2}\,\frac{1}{q_0^2-m_\rho^2}\,,\\
\Pi_{V\,\mathit{med}}^{(1\,\text{n})} &=
  \frac{T^2 g \sigma_0^2 fq_0^2}{6f_\pi^2}\,
  \frac{h_1\sigma_0^2}{(q_0^2-m_{a_1}^2)(q_0^2-m_\rho^2)}\,,
\end{align}
and the sum of all contributions reads
\begin{align}
\Pi_{V\,\mathit{med}}
  &= \Pi_{V\,\mathit{med}}^{(1\,\text{h--l})}
     +2\,\Pi_{V\,\mathit{med}}^{(1\,\text{m--n})}\nonumber\\
  &= \frac{T^2}{6f_\pi^2}\Big(\frac{(fq_0^2)^2}{q_0^2-m_\rho^2}
     -\frac{(fq_0^2-g\sigma_0^2)^2}{q_0^2-m_{a_1}^2}-\sigma_0^2\Big)\,,
\label{mixingexplicit}
\end{align}
which is exactly the result expected from Eq.\ (\ref{mixing1loop}).

However, in the model for $\Pi_V$ and $\Pi_A$ we presented in Ref.\
\cite{UBW}, the correlators are not calculated in a strict one-loop
approximation. Instead, the Dyson series is summed, allowing for a
successful description of $\Pi_V$ in vacuum, but mixing all orders of
$\hbar$. Thus the dominant medium contributions to $\Pi_V$ are given
by diagrams similar to those shown in Fig.\ \ref{diagPi}, but with all
bare $\rho$ propagators $1/(q^2-m_\rho^2)$ replaced by dressed ones,
$G_{\rho\,\mathit{vac}}(q^2) =
1/[q^2-m_\rho^2-\Sigma_{\rho\,\mathit{vac}}(q^2)]$, and all point-like
$\gamma\rho$ vertices $f$ replaced by loop-corrected ones,
$F_{\gamma\rho\,\mathit{vac}}(q^2)$.

\begin{figure}
\includegraphics[width=7cm]{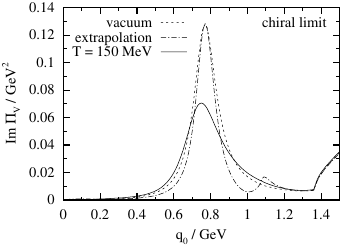}
\caption{The imaginary part of vector correlator, $\mathrm{Im}\, 
\Pi_V(q_0,0)$, in the chiral limit as function of energy, in vacuum
(dashed line), as an extrapolation of the low-temperature behavior
calculated for $T = 25$ MeV to $T = 150$ MeV (dash-dotted line), and
at $T = 150$ MeV (solid line).
\label{PiVx}}
\end{figure}

In this approximation one can see that, to order $T^2$, the imaginary
part of $\Pi_V$ (i.e., the spectral function) decreases in the
neighborhood of $m_\rho$, but exactly at $m_\rho$ it remains constant,
i.e., the strength of the $\rho$ meson peak is reduced by reducing its
width and not its height. In addition, similar to Eq.\
(\ref{mixingexplicit}), a peak $\propto\delta(q_0^2-m_{a_1}^2)$ is
generated, but the coefficient is different from that in Eq.\
(\ref{mixingexplicit}). For demonstration, we have computed
$\mathrm{Im}\,\Pi_V$ at $T = 25$ MeV and extrapolated the medium
contribution to $T=150$ MeV as if there were no effects beyond order
$T^2$. The result can be seen in Fig.\ \ref{PiVx} together with
$\mathrm{Im}\,\Pi_{V\,\mathit{vac}}$.
\begin{figure}
\includegraphics[width=7cm]{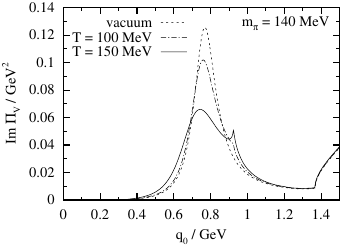}
\caption{The imaginary part of vector correlator, $\mathrm{Im}\,
\Pi_V(q_0,0)$, for a realistic parameter set as function of energy, at
various temperatures: $T = 0$ (dashed line), $T = 100$ MeV
(dash-dotted line), and $T = 150$ MeV (solid line).
\label{PiVa}}
\end{figure}
The third curve in Fig.\ \ref{PiVx} shows $\mathrm{Im}\,\Pi_V$
calculated at $T=150$ MeV. Obviously this looks completely different
from the extrapolation of the low-temperature behavior. The peak at
$q_0 = m_{a_1}$ is washed out by the thermal motion of the pions,
which is so strong that the $\pi a_1$ loop [diagram (d)] generates an
imaginary part not only at $q_0\approx m_{a_1}$, but also at
$q_0\approx m_\rho$. Another important contribution is given by the
two-pion loop, although it is of the order $T^4$ as discussed before
Eq.\ (\ref{Sigmarhob}). These two effects are the reasons for the
broadening of the $\rho$-meson peak visible in Fig.\ \ref{PiVx}.

To conclude this letter, we display in Fig.\ \ref{PiVa} some results
for the vector correlator obtained with the realistic parameter set,
i.e., $m_\pi = 140$ MeV.
In this case, of course, the mixing in the form of Eq.\ (\ref{mixing})
cannot occur, since all temperature effects are suppressed by
$e^{-m_\pi/T}$, and the axial correlator mixed into the vector
correlator at energy $q_0$ must be taken at energies below $q_0-m_\pi$
or above $q_0+m_\pi$. Indeed, with increasing temperature, in addition
to the broadening of the $\rho$-meson peak, a sharp peak at
$q_0 = m_{a_1}-m_\pi$ shows up in $\mathrm{Im}\,\Pi_V$, which can be
interpreted in this sense. Unlike the peak at $q_0 = m_{a_1}$ in the chiral
limit, this peak is not smeared by the thermal motion of the pions,
because it results from a kink in $\Sigma_\rho$ at the threshold $q_0
= m_{a_1}-m_\pi$, which persists to arbitrary temperatures. The
existence of this threshold, and therefore of the peak in
$\mathrm{Im}\,\Pi_V$, is of course an artifact of our approximation,
in which the $a_1$-mesons inside the loops have no width. In a
more realistic model this reminiscence of the Dey-Eletsky-Ioffe mixing
would reduce to a smooth enhancement of $\mathrm{Im}\,\Pi_V$ above
the $\rho$-meson peak. In order to see this, an approximation scheme,
in which also the particles inside the loops are dressed
self-consistently, is desirable \cite{vanHees}.

In order to draw quantitative conclusions for observables which are
directly accessible by heavy-ion experiments (e.g., dilepton rates),
also the effects of finite three-momentum $\bm{q}$, and, more
importantly, of finite baryon density, must be included. The former is
straightforward, but the inclusion of baryonic effects in a chirally
symmetric way is a very difficult task and is postponed to future
work.

\begin{acknowledgments}
One of us (MU) acknowledges support from the Alexander von Humboldt
foundation and would like to thank Prof. G. Chanfray for the
hospitality at the IPN Lyon. This work was supported in part by
the BMBF.
\end{acknowledgments} 

\end{document}